\documentclass[amsmath,amssymb,aps,pre,reprint,showpacs]{revtex4-1}     
\usepackage{bm}
\usepackage{dcolumn}
\usepackage{graphicx}

\usepackage[caption=false]{subfig}

\begin{document}
\preprint{APS/123-QED}

\title{Localized states in periodically forced systems}
\author{Punit Gandhi}
 \email{punit\_gandhi@berkeley.edu}
\affiliation{Department of Physics, University of California, Berkeley CA 94720, USA}
\author{C\'edric Beaume}
\affiliation{Department of Aeronautics, Imperial College London, London SW7 2AZ, UK}
\author{Edgar Knobloch}
\affiliation{Department of Physics, University of California, Berkeley CA 94720, USA}
\date{\today}

\begin{abstract}
The theory of stationary spatially localized patterns in dissipative systems driven by time-independent forcing is well developed. With time-periodic forcing related but time-dependent structures may result. These may consist of breathing localized patterns, or states that grow for part of the cycle via nucleation of new wavelengths of the pattern followed by wavelength annihilation during the remainder of the cycle. These two competing processes lead to a complex phase diagram whose structure is a consequence of a series of resonances between the nucleation time and the forcing period. The resulting diagram is computed for the periodically forced quadratic-cubic Swift--Hohenberg equation, and its details interpreted in terms of the properties of the depinning transition for the fronts bounding the localized state on either side. The results are expected to shed light on localized states in a large variety of periodically driven systems.


\end{abstract}

\maketitle


Spatially localized structures arise in a number of systems of interest in physics, chemistry and biology \cite{champneys1998homoclinic,Richter05,Batiste2006spatially,knobloch2008spatially,schneider2010snakes,Beaume2013convectons}. Such states often consist of a steady spatial pattern embedded in a homogeneous background. The theory of these states is well developed, at least in one spatial dimension, where the spatial coordinate $x$ can be employed as a time-like variable to describe solutions on the real line that evolve away from the homogenous state as $x$ increases from $-\infty$ before returning to it as $x\to\infty$ \cite{balmforth1995solitary,woods1999heteroclinic,burke2006,burke2007stability,houghton2009,Kozyreff09,makrides14}. In many cases, however, the localized states may be embedded in a fluctuating background \cite{schapers2000interaction,Prigent02,Barkley05} or the system may be subject to time-dependent forcing \cite{umbanhowar1996localized,Lioubashevski99,Rucklidge2009design,alnahdi2013localized}, situations to which the current understanding does not apply. Of particular interest is the study of vegetation patterns that arise in semi-arid regions \cite{tlidi2008vegetation,meron2012pattern}. Such systems are often bistable between a bare soil state and a vegetation state, and exhibit localized structures. In models such patterns may gradually shrink in extent or collapse homogeneously, depending on the level of precipitation \cite{zelnik13}. In this Letter, we study the processes governing the growth or decay of such patterns in systems subject to time-periodic forcing, eg., seasonal variation in growing conditions (precipitation, for instance) and map out the location in parameter space where localized states persist or decay. The intricate structure we find is a consequence of resonances between the growth timescale and the forcing period. 

In the absence of periodic forcing the localized states in these systems are found within a pinning or snaking region in parameter space \cite{pomeau} whose structure is captured in detail by the Swift--Hohenberg equation with competing nonlinear terms. This is so for shear flows~\cite{Schneider2010localized}, convection~\cite{mercader2011convectons}, optical systems~\cite{firth2007proposed}, and even models of crime hotspots~\cite{short2010nonlinear,lloyd2013localised}. Related equations are used to model localized vegetation patches \cite{meron2012pattern,zelnik13}. As a result the Swift--Hohenberg equation has become the model of choice for studying spatial localization in different settings. We therefore adopt a model of this type to study the properties of localized states in systems with spatially homogeneous but temporally periodic forcing:
\begin{equation}
\partial_t u = r(t)u - \left(1 + \partial_x^2\right)^2 u + b u^2 - u^3.
\label{she}
\end{equation}
When $r$ is a constant and $b>\sqrt{27/38}$ this equation exhibits bistability between a stable homogeneous state $u_h(x) \equiv 0$ and a stable spatially periodic state $u_p(x)$. Within the bistability region one finds an infinite number of different stable localized states with symmetry under $x\to - x$ and either maxima (hereafter $L_0$ states) or minima (hereafter $L_{\pi}$ states) at $x=0$. These are located on two distinct branches within a pinning or snaking region that straddles the so-called Maxwell point $r\equiv r_M$, where $u_h$ and $u_p$ have the same free energy $E$ \cite{woods1999heteroclinic,burke2006}. For example, when $b=1.8$ bistability is present in $r_p<r<0$, where $r=r_p\approx -0.3744$ corresponds to a fold on the $u_p$ branch, while the localized states are found in the interval $r_- <r<r_+$, where $r_- \approx -0.3390$ and $r_+ \approx -0.2593$ \cite{burke2006}. When $r=r(t)$, Eq.~(\ref{she}) no longer has a Lyapunov structure and a free energy cannot be defined, but we can define an effective Maxwell point $r\equiv {\bar r}_M$ using the definition ${\bar E}=0$, where ${\bar E}=\tfrac{1}{T}\int_{0}^{T} E \,dt$ and $T$ is the oscillation period of $r(t)$.
\begin{figure*}
\begin{center}
\includegraphics[width=15cm]{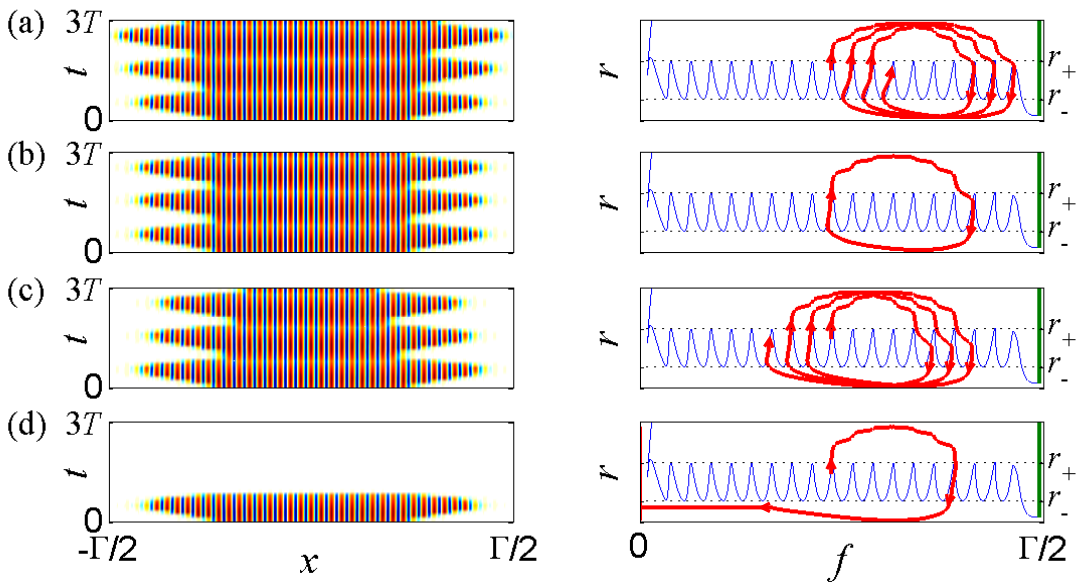}
\end{center}
\caption{The four different types of dynamics obtained from Eq.~(\ref{she}): growth by depinning at $r_0=-0.278$ (a), periodic orbit at $r_0 = 0.279$ (b), shrinkage by depinning at $r_0 = -0.280$ (c), and amplitude decay at $r_0=-0.281$ (d). Each solution is represented in a space-time plot (left) and as a trajectory superposed on a bifurcation diagram representing the forcing $r$ as a function of the location of the front $f$ (right). The bifurcation diagrams represent spatially localized ($L_0$, blue) and spatially periodic ($u_p$, green) states of the static system $r(t) \equiv r$. Parameters are $T=300$, $\rho = 0.1$ and $b=1.8$.
}
\label{samplesol}
\end{figure*}
To study the effect of parameter oscillation on the localized structures within the pinning region, we take
\begin{equation}
r(t) = r_0 + \rho \sin \left(\frac{2\pi t}{T} \right),
\label{timedep}
\end{equation}
with $r_-<r_0<r_+$. We choose the oscillation amplitude $\rho= 0.1 > (r_+ - r_-)/2 \approx 0.04$ so that the system exits the pinning region on either side during each cycle. When this is the case and $r>r_+$ the fronts connecting the localized state to $u_h$ temporarily depin and the structure grows by nucleating additional wavelengths of the spatial pattern. When $r<r_-$ the fronts also depin but this time the structure starts to shrink as the homogeneous state starts to invade the periodic state. For $r<r_p$ this process is overwhelmed by an overall decay of the whole structure.  In the limit of fast oscillations leading order asymptotics predict that $\bar{r}_M$ behaves as an effective Maxwell point of the averaged system, with a region of bistability and a pinning region that shrink as $T$ increases. However, for larger $T$ this is no longer the case and the dynamics is instead organized by a series of resonances between the cycle period and the time required to nucleate/annihilate wavelengths of the pattern.

Our simulations employ a fourth-order time differencing scheme \cite{cox2002} coupled to a Fourier scheme in space. In all cases we use the stable spatially localized solutions $L_0$ of the $r \equiv r_0$ problem as initial conditions, and solve Eq.~(\ref{she}) on a periodic domain of length $\Gamma$ that is sufficiently large to avoid finite size effects, typically $\Gamma = 80\pi$, i.e., $40$ critical wavelengths. Similar results were obtained for $L_{\pi}$ initial conditions.


We characterize the length of a given localized solution in terms of the location of the fronts that connect it to the homogeneous state on either side: 
\begin{equation}
f = \frac{2}{||u||^2_{\Gamma/2}} \int_{0}^{\Gamma/2} xu^2\,dx, \quad ||u||^2_{\Gamma/2} = \int_{0}^{\Gamma/2} u^2\,dx.
\end{equation}
Thus the $L_0$ states for constant forcing are bounded by fronts at $x=\pm f$, where $f$ takes values near $(1+2n)\pi,\, n=0,1,2,\dots$.
With the forcing (\ref{timedep}), $0<T<\infty$, episodic depinning generates oscillations in the spatial extent of the localized state. The number of wavelengths lost and gained within a cycle depends on the parameters $r_0$, $\rho$ and $T$ and the relative balance determines whether the localized state ultimately grows through net nucleation (Fig.~\ref{samplesol}(a)), persists indefinitely (Fig.~\ref{samplesol}(b)), or decays by net annihilation (Fig.~\ref{samplesol}(c)). Amplitude decay begins to dominate the dynamics of the localized states outside of the region of bistability and, if enough time is spent with $r<r_p$, the amplitude of the structure may fall below a critical value from which it cannot recover (Fig.~\ref{samplesol}(d)). Figures~\ref{samplesol}(a)--(d) also show the corresponding projections on the $L_0$ bifurcation diagram \cite{burke2006} and reveal that within the pinning region the solution trajectory follows appropriate portions of the $L_0$ branch; nucleation events are triggered when it exits the pinning region into $r>r_+$ and these manifest themselves in sudden jumps in $f$. In contrast, the annihilation events occur in close succession and $f$ varies continuously during the decay phase of each cycle period.

We characterize the overall behavior by computing the change in $f$, averaged over a large number ($N_t\geq 10$) of periods $T$:
\begin{equation}
\langle \triangle f \rangle = \frac{f(t=t_0+N_tT) - f(t=t_0)}{N_t},
\end{equation}
where $t_0\geq T$ is taken large enough to bypass initial transients.
\begin{figure}[p]
\includegraphics[width=8cm, bb= 30 170 550 595,clip]{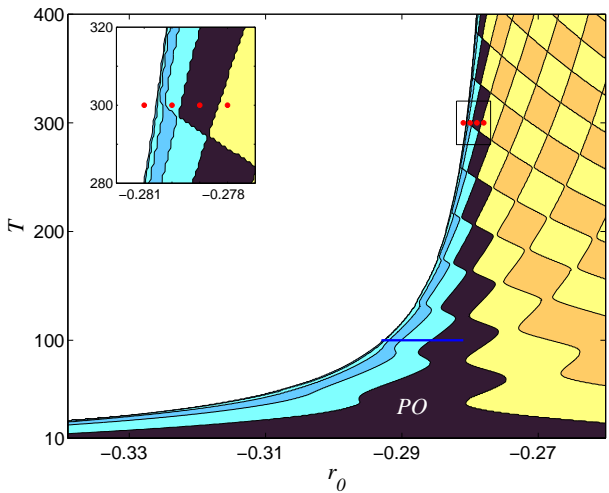}
\caption{Map of the different dynamics observed in the ($r_0,T$) plane for $\rho = 0.1$ and $b=1.8$. Periodic orbits exist in region $PO$ (Fig.~\ref{samplesol}(b)). The regions to the right (left) of $PO$ consist, in order, of growing (decaying) solutions where the pattern experiences net growth (decay) by $1,2,\dots$ wavelengths on either side per cycle as exemplified in Fig.~\ref{samplesol}(a) (Fig.~\ref{samplesol}(c)). The white region indicates parameter values at which the amplitude of the localized pattern decays within one cycle independently of its original size (Fig.~\ref{samplesol}(d)). The figure is plotted over the pinning region $r_-<r_0<r_+$ for $\rho=0$. The red dots refer to solutions shown in Fig.~\ref{samplesol} while the blue line indicates the parameter values investigated in Fig.~\ref{cliff}.
}
\label{wholediag}
\end{figure}

\begin{figure}
\begin{center}
\includegraphics[width=7.5cm, bb= 25 175 575 600,clip]{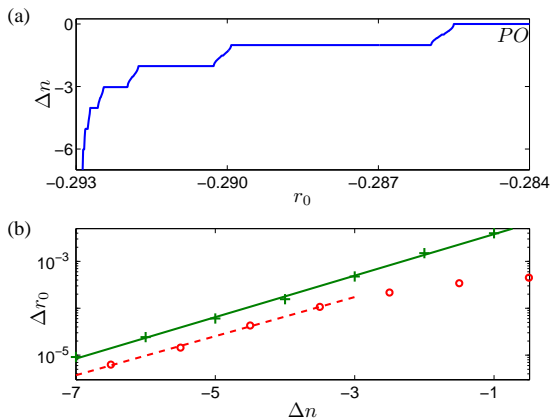}
\caption{(a) Average change in the number of wavelengths $\Delta n = \langle \triangle f \rangle / 2 \pi$  on each side of the localized structure per cycle of the forcing as a function of $r_0$ for $T=100$, $\rho = 0.1$ and $b=1.8$ (blue line segment in Fig.~\ref{wholediag}). (b) Length $\triangle r$ of the plateaus (green crosses, plotted at integer values of $\Delta n$) and of the transition regions between plateaus (red circles, plotted at $\Delta n+0.5$ for a transition between plateaus $\Delta n$ and $\Delta n+1$) as functions of $\Delta n$. The best fit lines, $\triangle r_P(\Delta n)=1.1\times 10^{-2}\exp(1.02 \Delta n)$ for the plateaus and $\triangle r_T(\Delta n) = 3.3\times 10^{-3}\exp(0.96 \Delta n)$ for the transition zones, indicate that both shrink exponentially as $|\Delta n|$ increases.}
\label{cliff}
\end{center}
\end{figure}

\begin{figure}
\begin{center}
\includegraphics[width=8cm, bb= 30 180  575 610,clip]{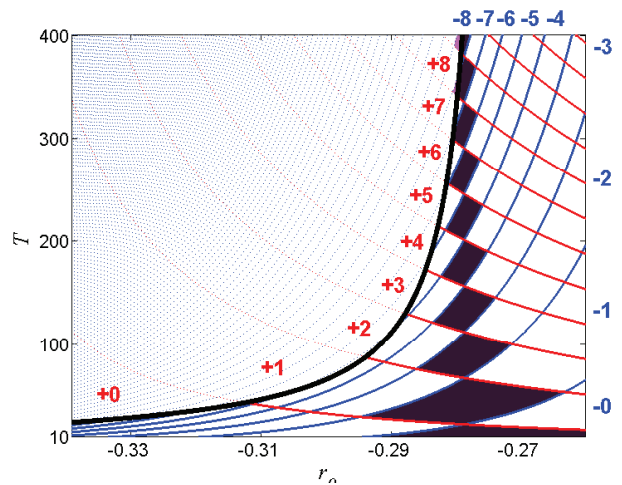}
\caption{Prediction in the $(r_0,T)$ plane of net decay (growth) of a spatially localized initial condition of Eq.~(\ref{she}) with time-periodic forcing (\ref{timedep}) obtained from the relation (\ref{asympt}) with $\rho = 0.1$ and $b=1.8$ and the numerical fit (\ref{nfit}). Positive (resp. negative) integers represent the change in the number of wavelengths on either side of the localized structure due to nucleation $[n_+]$ (resp. annihilation $[n_-]$) events during one forcing cycle; $\Delta n\equiv [n_+]+[n_-]=0$ in the dark region. The bold black line represents the predicted location of the cliff that marks the onset of amplitude collapse. The figure is plotted over the same $r_0$ interval as Fig.~\ref{wholediag}.}
\label{asymp}
\end{center}
\end{figure}
Simulations are run for $\rho = 0.1$, $r_- \le r_0 \le r_+$ and $10 \le T \le 400$ in increments of $10^{-4}$ in $r_0$ and $1$ in $T$. The net change $\langle \triangle f \rangle$ in position of the right front takes values near $0$, $\pm 2\pi$, $\pm 4\pi$, etc.; $\langle \triangle f \rangle=0$ indicates the presence of a periodic orbit in which $0,1,2,\dots$ wavelengths are added on either side of the structure and then annihilated within each cycle period. Nonzero values of $\langle \triangle f \rangle$ correspond to a net growth or decay per cycle of $1,2,\dots$ wavelengths on either side of the localized structure. The results are reported in Fig.~\ref{wholediag} where the boundaries between colored regions correspond to $\langle \triangle f \rangle = \pm  \pi, \pm  3\pi, \dots$. The central region of the $(r_0,T)$ parameter plane, labeled $PO$, corresponds to spatially localized, periodically breathing states (periodic orbits) characterized by a balance between the number of wavelengths nucleated in a cycle and the number annihilated. To the right of this region the number of wavelengths nucleated exceeds that destroyed and the structure gradually grows in length; the opposite is the case to the left of $PO$. Both processes accelerate with distance from this region, and in the white region in Fig.~\ref{wholediag} the structure collapses within one cycle regardless of its spatial extent. This is so even for domain-filling periodic states--a signature that the temporal forcing has effectively narrowed the parameter range of bistability for the system.  As $T$ increases the phase diagram exhibits a succession of {\it pinched zones}, where the region of periodic orbits shrinks dramatically, separated by {\it sweet spots}, where it expands again. These are centered around $r_0 \approx -0.29$ at low $T$ and slant to values slightly larger than $r_0 \approx -0.28$ as $T$ increases. A similar structure is observed in the regions of growing/decaying solutions. 
The variation in the pattern length to the left of the $PO$ region is reported in Fig.~\ref{cliff} for $T=100$. The figure reveals a series of plateaus corresponding to the loss of a fixed number of wavelengths per cycle. These are separated by thinner \textit{transition zones}, where the number of wavelengths lost per cycle is not an integer. For instance, in the transition zone between $PO$ and the first decay region one can find a region within which the pattern loses two wavelengths every three oscillation cycles. Figure \ref{cliff}(b) reports the size of the plateaus and of the transition zones as a function of the number of wavelengths lost per cycle and suggests that the left boundary of the $(r_0,T)$ region in Fig.~\ref{wholediag} corresponds to an accumulation point of exponentially decreasing intervals within which progressively more wavelengths are lost per oscillation cycle.  

Beyond this ``cliff'', the system spends enough time in $r<r_p$ that the solution reaches the trivial state via an overall amplitude decay within one cycle regardless of its spatial extent. The location of this cliff moves to higher values of $r_0$ as $T$ increases as a consequence of the increased time spent undergoing amplitude decay. In an infinite domain we expect the cliff to approach $r_0=r_p+\rho \approx -0.2744$ as $T\to \infty$, as this is the threshold for reaching $r<r_p$ during a forcing cycle.

To understand the structure seen in Fig.~\ref{wholediag} we examine the depinning process that takes place as soon as $r(t)$ exits the pinning region, allowing the stable periodic state to invade the homogeneous state ($r>r_+$) or vice versa ($r<r_-$). When $\rho\rightarrow (r_+-r_-)/2 \pm \delta$, $\delta\ll 1$, this process is slow and takes place on a $\mathcal{O}(\delta^{-1/2})$ timescale. This time is comparable to the time spent outside the pinning region when $T = \mathcal{O}(\delta^{-1})$. An asymptotic calculation of the depinning time \cite{Aranson00,burke2006} yields $T^{\text{dpn}}_{\pm} \approx (\alpha_{\pm} \delta_{\pm}^{1/2})^{-1}$, 
where $\delta_{\pm} = |r - r_{\pm}|$ and $\alpha_{+} \approx 0.1682 $, $\alpha_{-} \approx 0.2279$ (for $b=1.8$). Here and in the following, quantities with a $-$ (resp. $+$) subscript refer to decay (resp. nucleation) events on the left (resp. right) of the snaking region. This asymptotic result allows us to predict the change in the number of wavelengths on each side of the localized structure during an excursion of the parameter $r(t)$ outside the pinning region by computing
\begin{equation}
n_{\pm} =\pm \int_{\mathcal{T_{\pm}}} \frac{dt}{T^{\text{dpn}}_{\pm}(t)},
\label{asympt}
\end{equation}
where $\mathcal{T_-}$ (resp. $\mathcal{T_+}$) is the time spent on the left (resp. right) of the pinning region during a cycle. We assume that the system equilibrates to the nearest stable localized state during the traverse of the pinning region, erasing or completing unfinished nucleation/decay events, and so round $n_{\pm}$ to the closest integer at time $\mathcal{T_{\pm}}$, denoted by $[n_{\pm}]$. Although qualitatively correct, we obtained better accuracy with a fifth order fit to $T^{\text{dpn}}_{\pm}$.  To do so, we calculated the depinning time on the right (left) of the pinning region  from simulations of Eq.~(\ref{she}) as a function of the distance from $r_{+}$ ($r_{+}$), and fitted the results using a least squares method with the polynomial approximation
\begin{equation}
\left(T^{\text{dpn}}_{\pm}\right)^{-1} \approx \sum_{i=1}^5 \alpha_{i \pm} \delta_{\pm}^{i/2}.
\label{nfit}
\end{equation}
 The predictions of this procedure for different values of $r_0$ and $T$ are shown in the $(r_0,T)$ plane in Fig.~\ref{asymp}. The regions of constant $[n_+]$ (resp. $[n_-]$) exist between the red (resp. blue) lines in Fig.~\ref{asymp} and are labeled with red (resp. blue) integers. The sum $\Delta n= [n_+]+[n_-]$ indicates the net change in the number of wavelengths of the pattern on either side of the localized structure during one cycle of the forcing. The $PO$ region corresponds to locations where this sum vanishes. This region exhibits alternating pinching zones and sweet spots in excellent agreement with the results presented in Fig.~\ref{wholediag}, and this agreement extends to regions of net growth and shrinkage on either side of the $PO$ region; these predictions become more and more accurate in the adiabatic limit $T\to \infty$.

A separate calculation is required to identify the location of the cliff beyond which the solutions irrevocably collapse to the trivial state. We calculate this location from a fifth order numerical fit to the collapse time $T^{\text{col}}$ for periodic solutions $u_p$ of the constant $r$ problem with $r<r_p$. The result applies in the time-dependent problem whenever $r(t)$ falls below $r_p$ and the system spends sufficient time in this region for the solution to collapse, i.e., we integrate Eq.~(\ref{asympt}) with $T^{\text{dpn}}_{\pm}(t)$ replaced by $T^{\text{col}}(t)$ over the time interval spent in $r<r_p$, and use the condition $n=1/2$ to define the cliff. The result, shown using a thick black line (Fig.~\ref{asymp}), also agrees very well with that found in Fig.~\ref{wholediag}. The theory does not, however, capture the behavior of the exponentially compressed decay regions accumulating at the cliff that are present in the full problem (\ref{she})--(\ref{timedep}) or the complex transition zones between adjacent plateaus shown in Fig.~\ref{cliff}(a).

We have described the impact of parameter oscillations on spatially localized structures. Our results, obtained using the simplest model of such states, reveal that the parameter oscillation shrinks the existence region of stationary spatially localized solutions down to a connected series of sweet spots populated by spatially localized, periodically breathing states, separated by pinched zones. These states, which can model seasonal invasion and retreat of vegetation in simplest models of desertification \cite{tlidi2008vegetation,meron2012pattern}, are generated by temporary depinning of the fronts on either side, leading to an oscillation between the growth and decay of the structure over a forcing cycle. The presence of the sweet spots is a consequence of resonances between the natural growth rate of localized structures outside the pinning region and the forcing frequency, and it is these resonances that are responsible for the complex structure of the parameter plane of the system (Fig.~\ref{wholediag}). These resonances occur even in gradient systems such as Eq.~(\ref{she}) because the growth process is itself periodic in the frame of the depinned front. The features of this structure are predicted quantitatively within a theoretical framework based on the properties of the depinning time in the static system (Fig.~\ref{asymp}). This framework provides insight into the dynamics of the system, generalizes to other forms of the parameter oscillation $r(t)$ and provides a rationale for studying depinning of two-dimensional patterns \cite{Lloyd08,Kozyreff13}, including models of desertification \cite{zelnik13} with time-dependent forcing. 

Acknowledgement: This work was supported by National Science Foundation under Grants No. DMS-1211953 and CMMI-1233692. 

\bibliography{oscilPRL}

\end{document}